\documentclass{article}
\usepackage{spconf,amsmath,graphicx, color, soul, booktabs, makecell, float, amssymb}
\usepackage{enumitem}

\title{EMOVIS: EMOTION-OPTIMIZED IMAGE PROCESSING}
\name{Dor Barber, Rony Zatzarinni, Hava Matichin, Noam Levy}
\address{Intel Corporation}

\begin{document}
%
\maketitle

\begin{abstract}
In cinematography, visual attributes such as color grading, contrast, and brightness are manipulated to reinforce the emotional narrative of a scene. However, conventional Image Signal Processors (ISPs) prioritize scene fidelity, effectively neglecting this expressive dimension. To bring this cinematic capability to real-time camera pipelines during video capture, we introduce EMOVIS (EMotion-Optimized VISual processing). We establish a systematic mapping between a compact set of high-level emotional states (Happy, Calm, Angry, Sad) and low-level ISP controls—including color saturation, hue, brightness, local tone mapping, and sharpness—supported by a calibration user study with statistically significant effects across parameters. We propose a control framework that integrates these emotion-driven adjustments into standard ISP hardware without altering the underlying processing stages. Validation via blind A/B testing shows that viewers prefer the emotion-optimized rendering in 87\% of trials when the target emotion matches the scene context, indicating that emotion-aligned ISP control improves perceived suitability for expressive visual content.

\end{abstract}




\section{Introduction}
\label{sec:intro}
Images and videos are widely used to convey emotional messages, and practitioners intuitively adjust image processing parameters - such as color, contrast, tones, and others - to reinforce a desired emotional response. While this practice is well established in photography and filmmaking, its systematic treatment within computational image processing remains limited.

Prior work has established quantitative observations about specific image quality aspects and emotional responses. Studies have shown a consistent link between hue and saturation in affective perception \cite{valdez1994effects}, as well as brightness and contrast in emotional valence and arousal \cite{zhang2016feeling}. Additionally, a systematic review has consolidated a century of research on the emotional effects of color and brightness \cite{DoWeFeelColours2025}; these findings and others provide a scientific basis for treating emotion as a function of parameters that are directly manipulated within the image processing pipeline.

\begin{figure}
    \centering
    \includegraphics[width=1\linewidth]{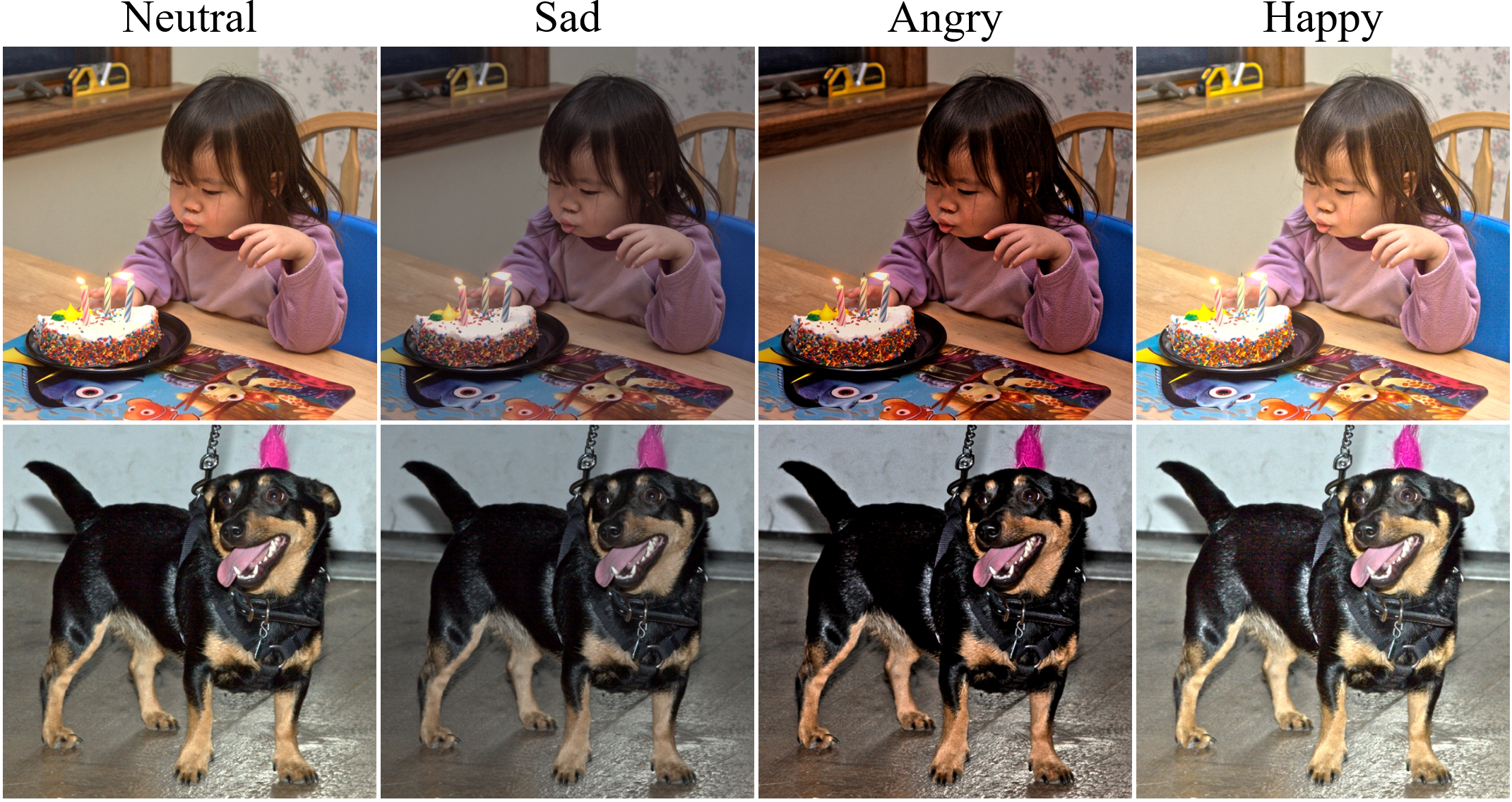}
    \caption{ISP configurations for different emotions: Neutral, Sad, Angry and Happy}
    \label{fig:intro}
\end{figure}

A substantial body of work focuses on detecting or classifying emotion in images and videos, along with the construction of corresponding datasets. While these efforts provide reliable representations of emotional context, they largely treat emotion as an output variable, rather than as a control signal for image processing itself \cite{borth2013large, yang2023emoset}.

Image quality is fundamentally subjective, motivating learning-based approaches such as NIMA \cite{talebi2018nima} that predict distributions of human ratings. More recent work also incorporates emotional context into the estimation of perceptual quality \cite{zhu2024emotion}; however, these methods focus on assessment rather than controlling the image formation process itself.

Emotion-aware image editing has been explored in some places; Ali and Ali \cite{ali2017automatic} show a framework for modifying the color distribution of images to achieve a specific emotional distribution; a more advanced method, using generative adversarial networks (GANs), is shown by Zhu et al.\ \cite{zhu2023emotional}. Although these methods demonstrate that image appearance can be altered to elicit target emotions, they operate offline in the image domain and do not address the constraints or real-time requirements of camera ISP pipelines. Beyond these limitations, achieving an ideal emotional resonance requires the holistic calibration of interdependent parameters—including saturation, brightness, and local contrast—rather than isolated adjustments.
Furthermore, post-processing approaches operate on developed, perceptual domain images with limited bit-depth. Attempts to aggressively manipulate color or contrast at this stage frequently result in quantization artifacts and the loss of highlight details due to clipping. By shifting this control upstream into the ISP, EMOVIS manipulates high-dynamic-range sensor data, preserving details that would otherwise be irretrievably lost.

The canonical ISP as described in the literature \cite{nakamura2017image}, exposes a structured set of parameters that govern the transformation from sensor data to the final image. While many parameters are fixed through calibrations (e.g. sensor noise models and color filter response;) others - such as tone mapping curves, color correction targets, and noise reduction strength - remain adjustable and directly affect perceived image quality. 


Recent work has demonstrated that ISP parameters can be algorithmically optimized using methods such as stochastic optimization, gradient-based learning, and reinforcement learning~\cite{adaptive_RL_isp_2024, rl_seq_ISP_2024}; in contrast, these approaches typically target task performance or generic perceptual metrics rather than emotional or contextual alignment.

Image quality evaluation is a long standing problem often relegated to experts; though the process clearly has a quantifiable aspect, much of what we call image quality is subjective or artistic \cite{understanding_artistic_2021}. Metrics designed to attempt to predict subjective image quality even using learned means often fail \cite{perceived_image_quality_2020} and the conclusion of the need for subjective evaluation remains. These limitations suggest that optimizing image quality requires explicit consideration of human perception, including both \textbf{cognitive} and \textbf{emotional} components. This observation motivates our approach of integrating emotional context directly into the image processing and evaluation loop.

We propose a framework that integrates the emotional context in both the evaluation and control of the ISP. We introduce a structured representation of the scene "vibe", map it to a subset of controllable ISP parameters suitable for real-time camera pipelines, and demonstrate through user studies that emotion-aware ISP control leads to higher perceived image preference compared to the emotion-agnostic baseline.

We establish our approach on psycho-visual traits and approach precise aspects of image processing, and then identify specific aspects of the ISP pipeline where those traits could be influenced with relative ease. We specifically propose changes within the framework of well established ISP algorithms to simplify the adoption of the approach in real-time processing.

To our knowledge, our approach is novel in combining an end to end approach of the entire image processing pipe, with the known effects of emotions on perception, and is shown with supporting user studies to directly influence the user preference of the image quality. Our contributions:
\begin{enumerate}[leftmargin=2.2em,labelsep=0.5em,itemsep=0pt,topsep=2pt,parsep=0pt]
\item Map emotions to ISP parameters, calibrated and supported by user studies.
\item Propose a framework to implement as part of the ISP.
\item Validate using a two-step user study with statistical analysis.
\end{enumerate}


\section{Methodology}
\label{sec:methodology}

The EMOVIS framework modifies specific aspects of image quality by inserting strategic modifiers into the standard signal processing chain. We define a six-dimensional control vector \(\mathbf{\alpha}\)  representing these modifiers:
\begin{equation}    
\mathbf{\alpha}=\begin{pmatrix}
    \alpha_S & \alpha_{YB} & \alpha_{RG} & \alpha_{LC} & \alpha_B & \alpha_P
    \label{eq:alphas}
\end{pmatrix}
\end{equation}

For each target emotion \(\mathbf{E}^i\), we determine an optimal parameter set \(\mathbf{\alpha}^i\)  
 to process the image. Our methodology proceeds in three stages: first, we construct a tunable ISP pipeline; second, we apply a user study to calibrate \(\mathbf{\alpha}^i\)  
 for specific emotions; and finally, we validate the efficacy of these tunings via statistical analysis and blind A/B testing.

\subsection{The EMOVIS ISP Pipeline}
\label{subsec:EmovisPipe}



To choose the subset of modifications that is best suited for our study, we relied heavily on previous studies that showed how the different aspects of image quality correspond to emotional reaction, including red-green and yellow-blue bias \cite{valdez1994effects, DoWeFeelColours2025}, brightness, color saturation \cite{zhang2016feeling}, and sharpness \cite{cho2011effect}. To these, we added local contrast manipulation; we are not aware of any prior work that directly linked it to emotional response, yet we incorporated this parameter since psychophysical evidence indicates that modulation of local contrast levels drives visual salience and physiological arousal \cite{Miklosi2023}.
\begin{figure}
    \centering
    \includegraphics[width=1\linewidth]{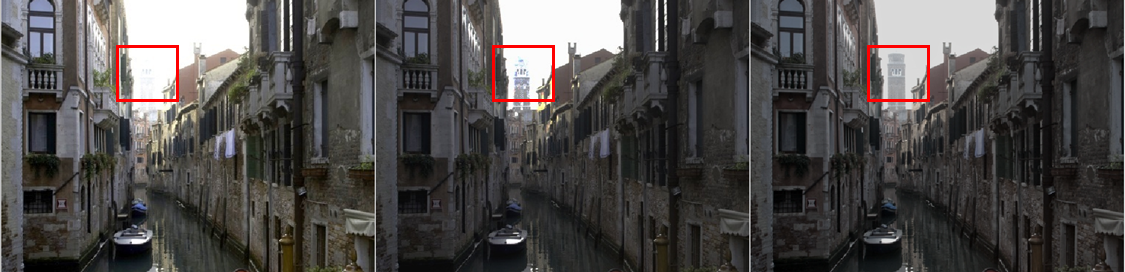}
    \caption{Importance of ISP-domain processing. Left: neutral image (FiveK processed image). Middle: post-processing baseline, generated by inverse tone mapping of the FiveK processed image to approximate linear input, followed by EMOVIS controls on processed 8-bit data to \textit{sad}. Right: EMOVIS full ISP processing from 16-bit RAW to \textit{sad}}
    \label{fig:ISPvsPostExample}
\end{figure}

Integrating controls directly into the ISP ensures superior fidelity compared to post-processing. While post-processing operates on tone-mapped, 8-bit sRGB images—often causing clipping and gamut artifacts—EMOVIS manipulates high-precision (12-14 bit) linear data. This preserves texture in saturated regions where post-processing fails (Figure~\ref{fig:ISPvsPostExample}). The following describes how these controls were implemented into an otherwise canonical ISP pipeline \cite{nakamura2017image}.

The \textbf{color saturation} is scaled by the factor $\alpha_S$. Let $\delta_S = 1+\alpha_S$ denote the target saturation gain, and $S=\|(C_R,C_B)\|_2$ denote the current chromaticity magnitude of a pixel. 
To prevent out-of-gamut artifacts when increasing saturation ($\alpha_S > 0$), the saturation gain is moderated based on the pixel's proximity to the gamut boundary:

\begin{equation}
\hat{\delta}_S = \min\left( \delta_S, \frac{S + 0.5(1-S)}{S + \epsilon} \right)
\end{equation}

The final chromaticity values are then updated using this moderated gain:
\begin{equation}
(C_R^S, C_B^S) = \hat{\delta}_S (C_R, C_B).
\end{equation}

We apply emotional \textbf{color biases} along the Red-Green ($\alpha_{RG}$) and Yellow-Blue ($\alpha_{YB}$) axes. 
In order to preserve white balance, we implement a content-aware masking strategy. We first estimate the saturation level of each pixel to derive a weight $w \in [0,1]$. This weight effectively acts as a mask, identifying colorful regions while ignoring neutral tones. 

Let $\mathbf{I} = [R, G, B]^T$ represent the input pixel vector. The weight $w$ is calculated using a quadratic formulation of the channel extrema:
\begin{equation}
    w = \left( \frac{\max(\mathbf{I}) - \min(\mathbf{I})}{\max(\mathbf{I}) + \epsilon} \right)^2
\end{equation}

Where $\epsilon$ is a small constant for numerical stability. The base term approximates the pixel's saturation; the squaring ensures a steeper attenuation as the pixel approaches gray, preserving neutral-tone stability.

The final tinted pixel, $\mathbf{I}'$, is computed by linear interpolation between the original pixel and a tinted version, governed by $w$:

\begin{equation}
    \mathbf{I}' = (1-w)\mathbf{I} + w(\mathbf{M}\mathbf{I})
\end{equation}

Here, $\mathbf{M} = \text{diag}(m_R, m_G, m_B)$ is a diagonal scaling matrix containing the tint coefficients derived from our emotional parameters:

\begin{equation}
    \begin{bmatrix} m_R \\ m_G \\ m_B \end{bmatrix} = 
    \begin{bmatrix} 
    1 + \alpha_{RG} + \alpha_{YB} \\
    1 - \alpha_{RG} + \alpha_{YB} \\
    1 - |\alpha_{RG}| - \alpha_{YB}
    \end{bmatrix}
\end{equation}

This formulation enables effective hue shifts while maintaining relative luminance stability.

The tone mapping framework integrates \textbf{brightness} adjustment and \textbf{local contrast} control applied to the image luminance $Y$. First, the brightness modifier $\alpha_B$ modulates the target luminance in linear space. We calculate a correction power, $\beta$, to map the average luminance of the region of interest (e.g., face or frame average) to a target luminance $T$, scaled by $\alpha_B$:

\begin{equation}
    \beta = \frac{\log(T \cdot (1 + \alpha_B))}{\log(\text{avg}(Y_{ROI}))}
\end{equation}

The tone-mapped image luminance is then derived as $Y^G = Y^\beta$. Subsequently, following \cite{soni2020improved} , we utilize a guided filter to extract the base layer $B_y=F_{\textbf{guided}}(Y)$ from the linear domain $Y$. Denoting $C(\cdot)$ as the histogram operator (CLAHE) the final tone-mapped result $Y^{TM}$, with a local detail boost controlled by $\alpha_{LC}$, is:

\begin{equation}
    Y^{TM} = \left( 1 + (\zeta+\alpha_{LC}) \frac{Y-B_y}{B_y} \right) C(Y^{G})
\end{equation}

Where $\zeta$ is the contrast modifier derived from the baseline calibration. 
%
%
Finally, we modify the \textbf{image sharpness} using an unsharp-masking algorithm \cite{deng2010generalized}. The sharpening gain is the sum of the baseline neutral setting $p$ and the emotion-specific modifier $\alpha_P$. The sharpened image luminance $Y^S$ is computed as:

\begin{equation}
    Y^S = Y + (p+\alpha_P) (Y - G_\sigma (Y)) \cdot M(Y)
    \label{eq:sharpening}
\end{equation}

Where $G_\sigma$ denotes a Gaussian blur kernel and $M(Y)$ is an overshoot protection mask.

Other ISP algorithm controls, including temporal and spatial denoising power, \(\Gamma\)-correction, white balance preferences, and more were left out of the study, as we focused on parameters most supported by prior emotion-perception evidence and widely exposed in hardware ISPs, and to simplify the following user studies.

\subsection{Calibrating For Emotions}
\label{subsec:calibrating}
\begin{figure}
    \centering
    \includegraphics[width=.75\linewidth]{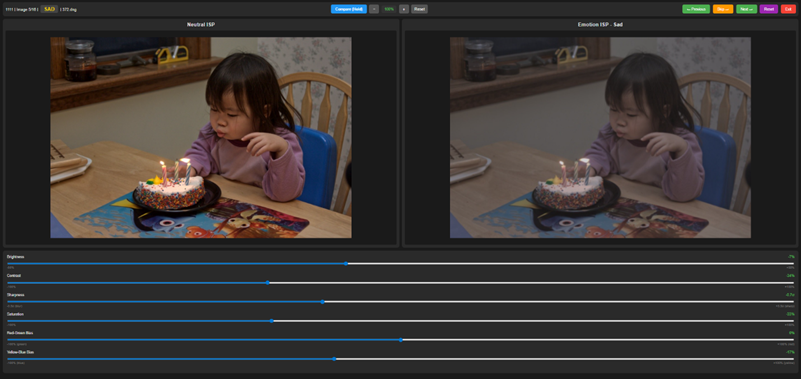}
    \caption{Screenshot of the tool developed for the calibration study as described in Section~\ref{subsec:calibrating}}
    \label{fig:theTools}
\end{figure}

We integrated EMOVIS ISP parameter tuning into an interactive tool (shown in Figure~\ref{fig:theTools}) and conducted a user study aimed at both calibrating emotion-specific ISP parameter adjustments and assessing the consistency of these preferences across participants, scenes, and images.

To select a compact and interpretable set of emotions for this study, we adopt the circumplex model of affect described by Mehrabian and Russell~\cite{mehrabian1974approach}, which represents emotional states in a continuous Pleasure--Arousal space (here treated as equivalent to Valence--Arousal). For the purposes of ISP control and user evaluation, this space is discretized into four high-level emotion super-classes: \textit{Happy}, \textit{Calm}, \textit{Angry}, and \textit{Sad}. These categories broadly correspond to regions of high or low valence and arousal and encompass related affective states (e.g., Happy: joyful, amused; Calm: relaxed, serene; Angry: hostile, tense; Sad: lonely, subdued). This discretization balances expressive coverage with experimental simplicity and aligns with the emotion categories used throughout the calibration and validation studies.

Fourteen participants took part in the calibration study. Each participant adjusted ISP parameters for a randomly selected subset of images drawn from the \textit{FiveK} dataset \cite{fivek}, resulting in a total of $N = 205$ calibrated samples. The study followed a within-subject design, with both images and target emotions randomized across trials. Participants were instructed: \textit{Use the sliders to select a visual appearance for the image that best matches the target emotion.}

The resulting parameter distributions were analyzed using repeated-measures ANOVA to assess the statistical significance of emotion-driven parameter differences. The calibrated parameter values are summarized in Table~\ref{tab:CalibrationResultsValues}, and the statistical analysis is reported in Section~\ref{sec:results}.

We note that for $\alpha_S$, we got an inverted behavior than expected \cite{DoWeFeelColours2025}; we explain this by the additional control of other tone/color parameters ($\alpha_B$, $\alpha_{LC}$, $\alpha_{RG}$), which allowed more expressive freedom to the tested users.

\subsection{Validation via A/B testing}
\label{subsec:validation}
Once the parameters of the EMOVIS ISP have been selected, we may now proceed to test the full pipeline. The method chosen was A/B testing.

Twenty four participants were selected, each randomly shown 16 videos from VEATIC \cite{ren2024veatic}, for a total of \(N=384\) samples. On half of the tests, one of the sides showed the neutral processing, while the other showed the processing matching the video sequence emotion as derived from the dataset labels (see Section~\ref{sec:dataset}). On the other half of the set, the users were shown the Neutral vs. one of the wrong emotion processing (different than the label.) Each user was asked the question: \textit{As a movie director or content creator, which video result would you prefer to use?}

\begin{table}
    \centering
    \begin{tabular}{ccccc}\toprule
         & Angry& Calm & Happy & Sad\\\midrule
        \(\alpha_S\)   & \(0.15\) & \(0.0\) & \(0.2\) & \(-0.18\) \\
        \(\alpha_B\)   &  \(-0.08\)& \(0.0\) & \(0.19\)& \(-0.09\) \\
        \(\alpha_{LC}\)   &  \(0.32\) & \(0.0\) & \(0.14\) & \(-0.02\) \\
        \(\alpha_{RG}\)   &  \(0.19\) & \(0.0\) & \(0.0\) & \(0.0\) \\
        \(\alpha_P\)   &  \(0.7\) & \(-0.2\) & \(0.0\) & \(0.0\) \\        
        \(\alpha_{YB}\)   &  \(0.0\) & \(0.0\) & \(0.0\) & \(-0.1\)\\ \bottomrule
    \end{tabular}
    \caption{Derived values of calibration parameters  based on the calibration study}
    \label{tab:CalibrationResultsValues}
\end{table}


\section{Datasets}
\label{sec:dataset}
To perform the user study and calibration, we chose the FiveK dataset \cite{fivek} of DNG raw images. We manually removed images with clear emotional bias (such as a person smiling). A raw camera format is required since the ISP needs the high precision to correctly convert the image without artifacts, and the FiveK dataset includes a wide range of image types including objects, persons, nature and still life. All the images came with optimally processed versions, which is our 'neutral' baseline.

For A/B testing as described in Section~\ref{subsec:validation}, we selected the VEATIC \cite{ren2024veatic} dataset consisting of minute-long clips; selecting short videos as opposed to still images was important to allow the user to fully comprehend the emotional backdrop of the scene before selecting a preference. Each clip includes a vector of \textbf{valence} \((V)\) and \textbf{arousal} \((A)\) values, with values centered around 0, negative being low and positive being high. Following the approach of \cite{mehrabian1974approach}, using the average \(V\) and \(A\) over the clip, we map the space as shown in Table~\ref{tab:VEATIC_Labels}. All the clips sampled had clearly negative or positive value for both metrics, so there was no need to handle border cases.

\begin{table}
    \centering
    \begin{tabular}{ccccc}\toprule
         & Happy & Calm & Angry & Sad \\\midrule
        \(\mathrm{sign}(V)\)  & \(+\) & \(+\) & \(-\)& \(-\) \\
        \(\mathrm{sign}(A)\)   & \(+\) & \(-\) & \(+\) &\(-\)  \\ \bottomrule
    \end{tabular}
    \caption{Emotion quadrants from Valence-Arousal}
    \label{tab:VEATIC_Labels}
\end{table}

Since VEATIC is a processed dataset, in order to apply ISP processing (which begins with a linear RAW file), we first applied \textit{inverse ISP} processing - restoring linear tones and sensor-domain colors; this introduced quantization artifacts to the processed result, and we address this limitation in our conclusions below. Since artifacts affected the processed results and our A/B test showed that subjects still preferred it with the appropriate emotion processing (see Section~\ref{sec:results} below), this limitation does not affect the conclusions of the study.


\section{Experimental Results}
\label{sec:results}

A repeated-measures ANOVA with emotion as a within-subject factor was conducted for each ISP parameter in the calibration study as described in Section~\ref{subsec:calibrating}. As summarized in Table~\ref{tab:CalibrationANOVA}, saturation, brightness, contrast, red–green bias, and sharpness all show statistically significant large effects (\(\eta^2 \geq 0.27\), \(p \leq 10^{-13}\)). Yellow–blue bias shows a statistically significant but small effect (\(\eta^2 =  0.056\)), suggesting that it plays a secondary role in emotion-driven ISP tuning. Effect sizes are interpreted following standard conventions \cite{cohen1988statistical}. Despite inter-participant variability, emotion-specific trends were consistent across images and participants, resulting in large effect sizes for most ISP parameters.

It is worth noting that the 'Calm' emotion resulted in parameter configurations nearly identical to the neutral baseline across most parameters, see Table 1. This aligns with the psychological definition of calm as a state of equilibrium with low arousal and neutral valence. Consequently, 'Calm' was excluded from the subsequent A/B validation phase, as the visual distinction from the neutral control was insufficient to yield meaningful preference data. The validation therefore focused on the three active emotional states: Happy, Sad, and Angry.

The A/B validation test confirms the efficacy of the pipeline for expressive content creation. As shown in Table~\ref{tab:ABtestResults}., when prompted to choose the best result for a \textbf{cinematic narrative}, viewers selected the emotion-optimized rendering in \textbf{87\%} of the trials where the ISP configuration matched the scene's emotional context.

Crucially, this preference is not merely a bias toward colorful or high-contrast images. When the ISP configuration contradicted the scene's context (e.g., applying a 'Happy' ISP to a 'Sad' scene), the preference for the processed image dropped to \textbf{24\%}. This stark difference indicates that EMOVIS successfully aligns the image rendering with the semantic content, improving the \textbf{perceived suitability} for storytelling rather than simply applying a generic aesthetic enhancement.

\begin{figure}
    \centering
    \includegraphics[width=1.0\linewidth]{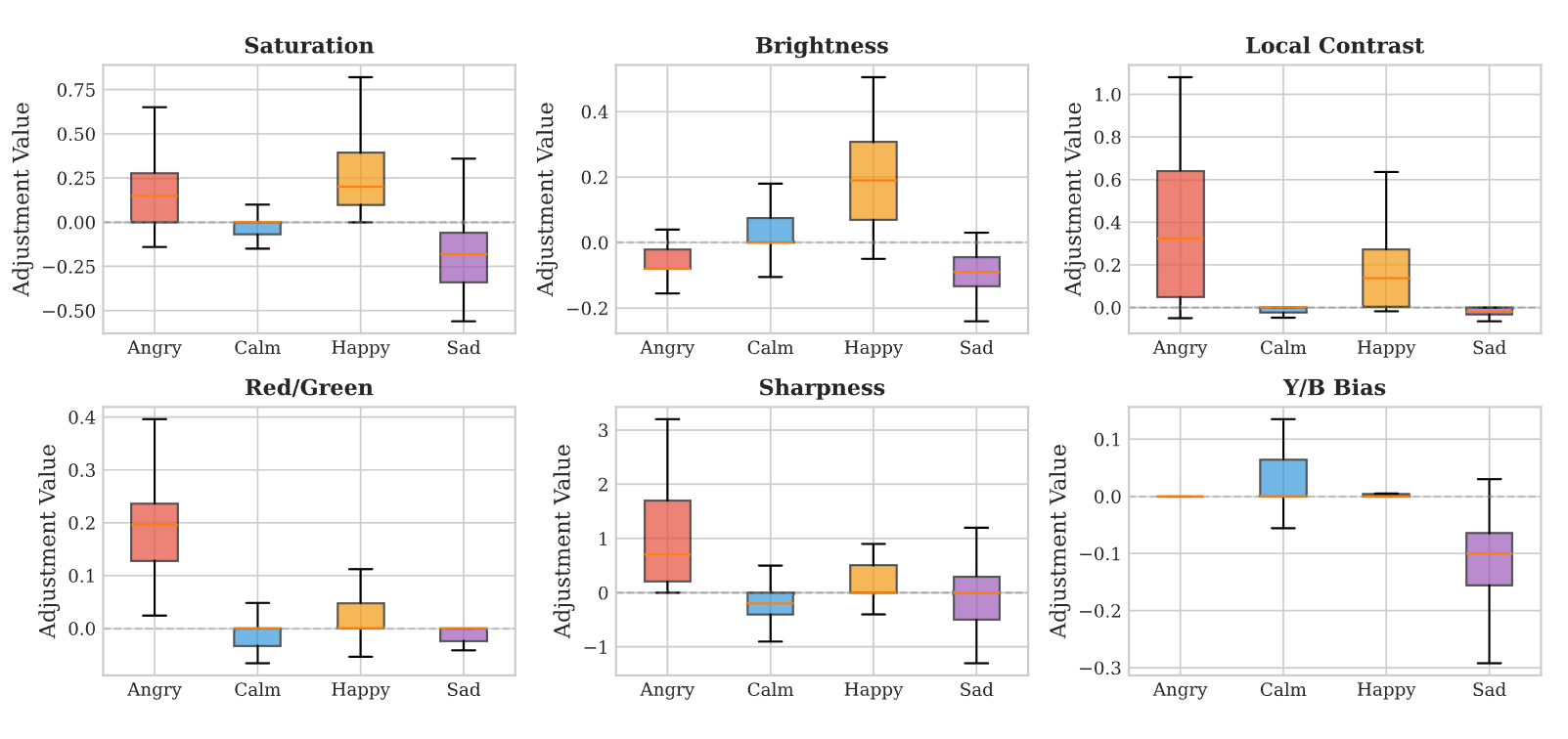}
    \caption{Box and whiskers plots of the calibration study results}
    \label{fig:CalibrationResultsBaW}
\end{figure}

\begin{table}[H]
    \centering
    \begin{tabular}{ccccc}\toprule
         & \(F_{stat}\) & \(p_{val}\) & \(\eta^2\) & Effect\\\midrule
        \(\alpha_S\)   & \(48.7\) & \(1.1 \times 10^{-23}\) & \(0.42\) & Large \\
        \(\alpha_B\)   & \(61.8\) & \(2.4 \times 10^{-28}\) & \(0.48\) & Large \\
        \(\alpha_{LC}\)   & \(49.0\) & \(8.0 \times 10^{-24}\) & \(0.42\) & Large \\
        \(\alpha_{RG}\)   & \(28.6\) & \(1.9 \times 10^{-15}\) & \(0.30\) & Large \\
        \(\alpha_P\)   & \(24.7\) & \(1.2 \times 10^{-13}\) & \(0.27\) & Large \\        
        \(\alpha_{YB}\)   & \(4.0\) & \(0.9 \times 10^{-3}\) & \(0.06\) & Small \\ \bottomrule
    \end{tabular}
    \caption{Repeated-measures ANOVA for calibration study}
    \label{tab:CalibrationANOVA}
\end{table}

\begin{table}[H]
    \centering
    \begin{tabular}{ccc}\toprule
         &Prefer Emotion  & Prefer Neutral \\\midrule
         Correct Emotion&87\% & 13\%\\\midrule
         Wrong Emotion&24\%& 76\%\\ \bottomrule
    \end{tabular}
    \caption{Results of the A/B testing}
    \label{tab:ABtestResults}
\end{table}

\begin{figure}
    \centering
    \includegraphics[width=0.8\linewidth]{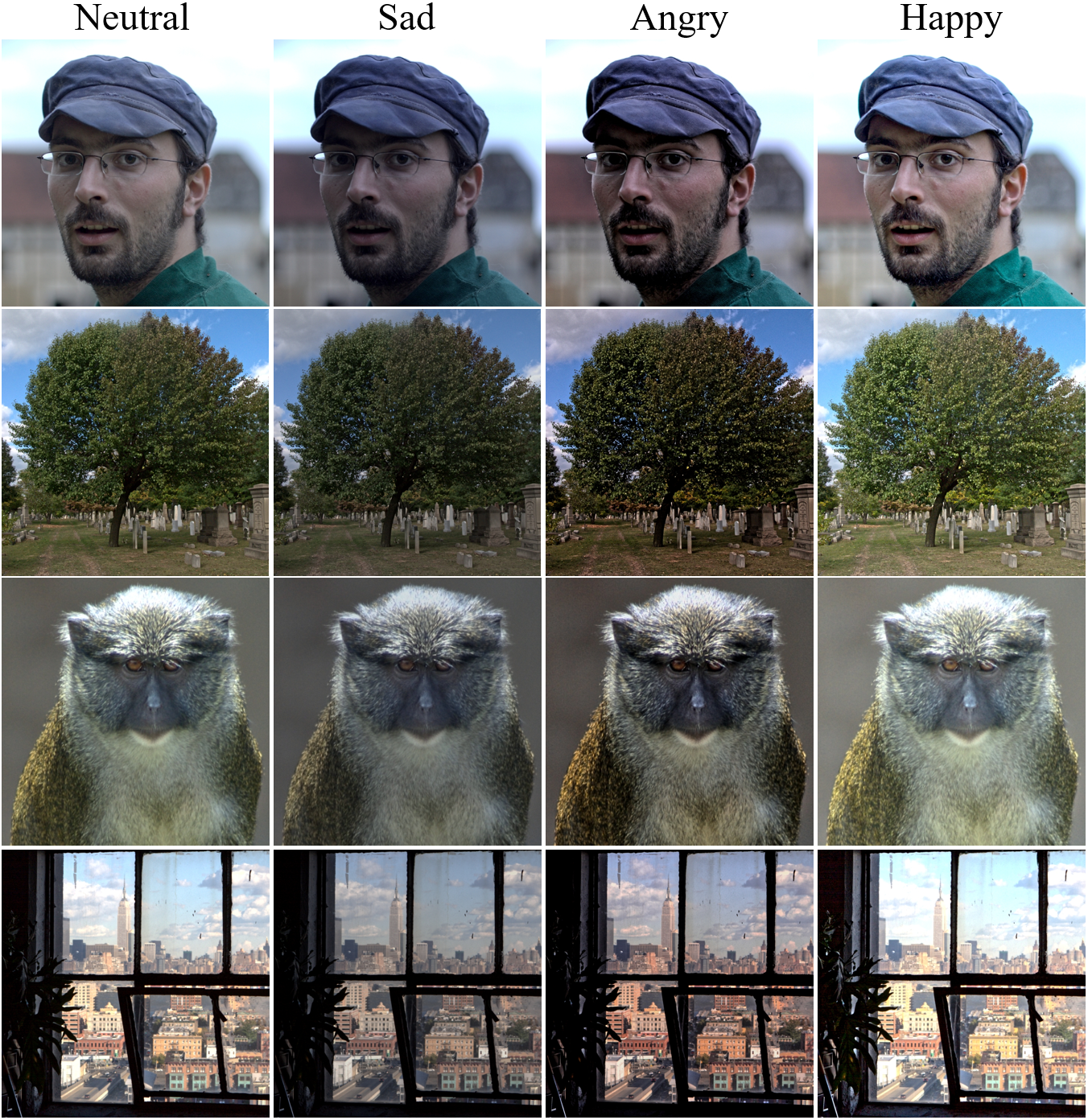}
    \caption{Results of our approach on several samples from the FiveK \cite{fivek} dataset}
    \label{fig:Results}
\end{figure}


\section{Conclusions}
\label{sec:conclusions}

We have shown that the EMOVIS ISP architecture, after being properly calibrated to 4 representative emotions, can cause the average user to prefer an emotionally-adapted ISP processing result over the neutral alternative, prompted to select images as content creators or film directors.

The results allow ISP developers to optimize their design to allow for the implementation of this framework and also open the door to a wider discussion on how the realm of subjective image quality and ISP tuning can be enhanced by incorporating the semantic understanding of the emotional backdrop of the image, video, or scene.

While providing important results, our approach is also limited: we have used a relatively small dataset for our experiments, explored only a limited set of discrete emotions, and did not evaluate a possible framework to incorporate a continuous emotion representation into the ISP processing; we have also not explored possible influences of emotional processing on different ISP algorithms and pipelines; finally, our validation was based on an inverse-ISP approach, which is inherently limited in the potential image quality. Future research can address these limitations.

Further work may also expand the framework to automatically detect emotion in the scene or image and integrate to one holistic approach; such a framework may then be validated by more extensive user studies similar to our approach.


\vfill\pagebreak

\bibliographystyle{IEEEbib}
\bibliography{strings,refs}

\end{document}